\documentclass[onecollarge,final,sort&compress]{svjour2}
\usepackage[numbers]{natbib}
\journalname{Few Body Systems}
\usepackage{graphicx}
\usepackage{slashed}
\usepackage{amstext}

\begin{document}

\title{First look at heavy-light mesons with a dressed quark-gluon vertex
\thanks{Presented by Mar\'ia G\'omez-Rocha at Light-Cone 2014, NC State University, Raleigh, USA, 26 - 30 May, 2014}
}

\author{Mar\'ia G\'omez-Rocha	\and
		Thomas Hilger	\and
        Andreas Krassnigg 
}

\institute{M. G\'omez-Rocha et al. \at
           University of Graz, Institute of Physics, NAWI Graz, A-8010 Graz, Austria\\
           \email{maria.gomez-rocha@uni-graz.at}  
}

\date{Received: \today}

\maketitle

\begin{abstract}
Following up on earlier work, we investigate possible effects of a dressed quark-gluon vertex
in heavy-light mesons. In particular, we study corrections to the popular rainbow-ladder truncation 
of the Dyson-Schwinger--Bethe-Salpeter equation system. We adopt a simple interaction kernel
which reduces the resulting set of coupled integral equations to a set of coupled algebraic
equations, which are solved numerically. In this way, we extend previous studies 
to quark-antiquark systems with unequal current-quark masses, at first for the pseudoscalar case,
and investigate the resulting set of problems and solutions.
We attempt to find patterns in -- as well as to quantify corrections to -- the 
rainbow-ladder truncation. In addition, we open this approach to phenomenological predictions 
of the heavy quark symmetry.

\keywords{Heavy-light mesons \and Dyson-Schwinger Equations \and Bethe-Salpeter Equation}
\end{abstract}

\section{Motivation}
\label{sec:motivation}

In modern studies of Quantum Chromodynamics (QCD) meson states made of one light and one heavy quark
have attracted particular attention for a number of years. The reasons are manifold, e.\,g., the possibility
to construct a theoretical bridge between effective field theories valid in opposite regimes like chiral perturbation 
theory as well as perturbative nonrelativistic QCD and heavy-quark effective theory, respectively. 
Another challenging reason is the need to understand a system
determined by different scales of the order of $\Lambda_{\rm{QCD}}$ on one hand and the 
bottom-quark mass on the other. 

Among the most promising approaches that are able to deal with all of these requirements as well as
able to provide a solid and modern theoretical framework for bound-state studies are lattice-regularized QCD studies
and also methods of continuum quantum field theories. Our method of choice herein is the
Dyson-Schwinger-Bethe-Salpeter-equation (DSBSE) approach, which has been applied to various
problems of theoretical hadron physics in the past decades with increasing success and comprehension,
see, e.\,g., \cite{Bashir:2012fs,Cloet:2013jya} and references therein. 

Typically, sophisticated hadron studies in the DSBSE approach are numerical in nature and 
make use of a truncation of the infinite tower of coupled integral equations involved; see, e.\,g., 
\cite{Blank:2010bp,Krassnigg:2009zh,Fischer:2014xha} for the details of such a setup. A popular
variant is the so-called rainbow-ladder (RL) truncation, where the dressed quark-gluon vertex \cite{Alkofer:2008tt}
assumes a strongly simplified structure; it models the gluon-quark interaction and reduces the effort of
solving relevant equations to the Dyson-Schwinger equation (DSE) of the quark propagator, and subsequently
the quark-antiquark Bethe-Salpeter equation (BSE) with a suitable interaction kernel.

Despite its success, in particular for pseudoscalar mesons and their various properties
\cite{Maris:1997tm,Maris:2000sk,Holl:2004fr,Maris:2005tt,Holl:2005vu}, 
vector mesons \cite{Maris:1999nt,Bhagwat:2006pu}, 
the heavy-quark domain \cite{Maris:2006ea,Blank:2011ha}, 
and an extension to a quark-diquark and later three-quark formulation for baryons 
\cite{Eichmann:2007nn,Eichmann:2009qa,Sanchis-Alepuz:2011jn,Eichmann:2012mp,Sanchis-Alepuz:2013iia}, always see also references therein,
this truncation has appeared limited in certain respects. Typically such limitations appear
where the bare covariant (vector) structure left in the quark-gluon vertex in this truncation is expected to
be insufficient for the treatment of certain states, such as those identified as orbital-angular-momentum
excitations \cite{Fischer:2008wy,Fischer:2009jm}. 
A situation similar to the latter case is presented by a heavy-light meson state, since one
cannot expect the same simplicity as apparent in a pseudoscalar or vector meson ground state with equal-mass quarks.
While heavy-light meson states have been investigated in the DSBSE approach, cf.~\cite{Ivanov:1998ms} and references therein, 
the assumptions made there about the connection between the DSE and BSE kernels are different from 
the truncation scheme considered here and cannot easily be related to our present setup.

Herein, we investigate dressing effects in the quark-gluon vertex in a scheme developed earlier in the equal-mass-constituent 
case \cite{Bender:2002as,Bhagwat:2004hn} for a simple interaction model. This goes beyond an introduction
of trivial dressing factors accompanying the bare structure in the quark-gluon vertex. For the sake of brevity, 
we merely sketch the formalism and rely on illustrations to highlight the problems and possibilities
given by our study. We quantify dressing effects for select pseudoscalar states and present
further directions of our work. The calculations are performed in Euclidean space.

\section{Setup and interaction model}
\label{sec:setup}

Meson studies in the DSBSE approach have a long-standing successful record. In the context of the 
present setup it is always instructive to note that first simplified attempts at meson spectroscopy 
analogous to the ones undertaken today were performed already several decades ago \cite{Munczek:1983dx}.
In fact, the interaction model defined there is the basis for the investigation presented here.

In QCD the DSE for the quark propagator reads
\begin{equation}\label{eq:dee}
 S^{-1}(p)= i \slashed{p}+\mathbf{1}m_q + \int\frac{d^4q}{(2\pi)^4}g^2 D_{\mu\nu}(p-q)\frac{\lambda^a}{2}\gamma_\mu S(q)\Gamma^a_\nu(q;p)
\end{equation}
where $p$ and $m_q$ are the quark momentum and current mass; $S$, $D_{\mu\nu}$, and $\Gamma^a_\nu$ are the renormalized dressed 
quark propagator, gluon propagator, and quark-gluon vertex, respectively. The color structure of the interaction is encoded in the equation via
the superscript ${}^a$, $g$ is the strong coupling constant, and $S$ has the general Dirac structure of a fermion propagator.

In order to study mesons, one needs to solve the BSE that contains two dressed quark propagators, obtained
by solving Eq.~(\ref{eq:dee}), together with the quark-antiquark scattering kernel $K$ in the form
\begin{eqnarray}\label{eq:bse}
\Gamma(p;P)&=&\int^\Lambda_q\!\!\!\! K(p;q;P)\;S(q_+) \Gamma(q;P) S(q_-) \,,
\end{eqnarray}
where $\Gamma$ is the Bethe-Salpeter amplitude (BSA) and $\Lambda$ a regularization scale. The (anti-)quark momenta are given by
$q_+ = q+\eta P$ and $q_- = q- (1-\eta) P$; the momentum partitioning parameter $\eta
\in [0,1]$ is an in principle free parameter, which accounts for the arbitrariness 
in the definition of the relative quark-antiquark momentum $q$. In the case of equal-mass quarks,
the obvious (and most convenient) choice is $\eta=1/2$, but in the present work we have to pay some
attention to this particular parameter and its effects on the results, explained in detail below.

With $S$ precomputed, a choice for $K$ is needed to proceed with the solution of the BSE. Such a choice is made possible
via guiding restrictions by the axial-vector Ward-Takahashi identity (AVWTI), which guards the effects of chiral symmetry and
its dynamical breaking in any model description of hadrons that implements a consistent set of interactions. While the simplest
truncation of the DSBSE system to satisfy this constraint, the RL truncation, is well-known, steps beyond it have been
sought and developed with the goal of a nonperturbative systematic scheme \cite{Bender:2002as,Bhagwat:2004hn} that 
would enable numerical studies of increasing sophistication, such as 
\cite{Watson:2004jq,Watson:2004kd,Fischer:2005en,Matevosyan:2006bk,Matevosyan:2007cx,Fischer:2008wy,Fischer:2009jm,Williams:2014iea}. 
On a more general basis the AVWTI has been investigated
to find construction principles for the meson-BSE kernel for a given type of quark-gluon vertex 
\cite{Munczek:1994zz,Chang:2009zb,Heupel:2014ina}.

The setup employed here is a straight-forward adaptation of the one presented in \cite{Bhagwat:2004hn}, and we refer
the reader to this reference for more details regarding the formalism and a comprehensive set of results for both the 
quark-propagator dressing functions as well as results for pseudoscalar and vector mesons with equal-mass quarks.
Here we sketch only the most important ingredients of our study; more results and a detailed account of the involvement
due to the unequal masses of the constituents in the BSE will be published elsewhere \cite{GomezRocha:2014mn}.

In order to understand and discuss the results presented herein, two more ingredients are necessary, namely the interaction
model and the construction principle for the quark-gluon vertex.
To define the interaction in this model, we follow \cite{Munczek:1983dx,Bhagwat:2004hn} in replacing the gluon propagator by
\begin{equation} 
\label{mnmodel} g^2 \, D_{\mu\nu}(k) := 
\left(\delta_{\mu\nu} - \frac{k_\mu k_\nu}{k^2}\right) (2\pi)^4\, {\cal G}^2 \, \delta^4(k)\,. 
\end{equation} 
The parameter $\cal G$ defines the (constant) interaction strength, sets a scale in the computations, and is
determined a priori. All integrals collapse via the $\delta^4$-term and one is left with a coupled system of 
algebraic equations which are then solved numerically. Apart from this simplifying effect, one has to note as well
that also in the BSA several structures (more precisely those involving factors of the relative momentum) become
trivial, i.\,e., they vanish. As a result the manifest covariance of the BSA in this model is destroyed, which
immediately impacts on the role of the momentum partitioning parameter $\eta$: meson masses and other
results are no longer $\eta$-independent like in the case with all covariants taken into account. 

In practice
this means that the model assumptions oversimplify meson structure to the extent of partial loss of covariance and 
seem to have even more but indirect consequences such as: the lack of solutions for scalar and axial-vector meson states; 
the trivialization of the entire BSA for mesons with spin $J>1$ \cite{Krassnigg:2010mh,Fischer:2014xha}; the lack of
solutions for radially excited meson states; an unnatural connection of interaction strength and the resulting
amount of dynamical chiral symmetry breaking, in particular at finite temperatures \cite{Blank:2010bz}.
Nonetheless, the benefits outweigh these shortcomings in that one can probe dressing effects up to infinite
order in any given scheme as well as more complicated approaches and new ideas with unparalleled ease as compared to 
studies using more realistic interactions.

\begin{figure*}
\centering
  \includegraphics[width=0.42\textwidth]{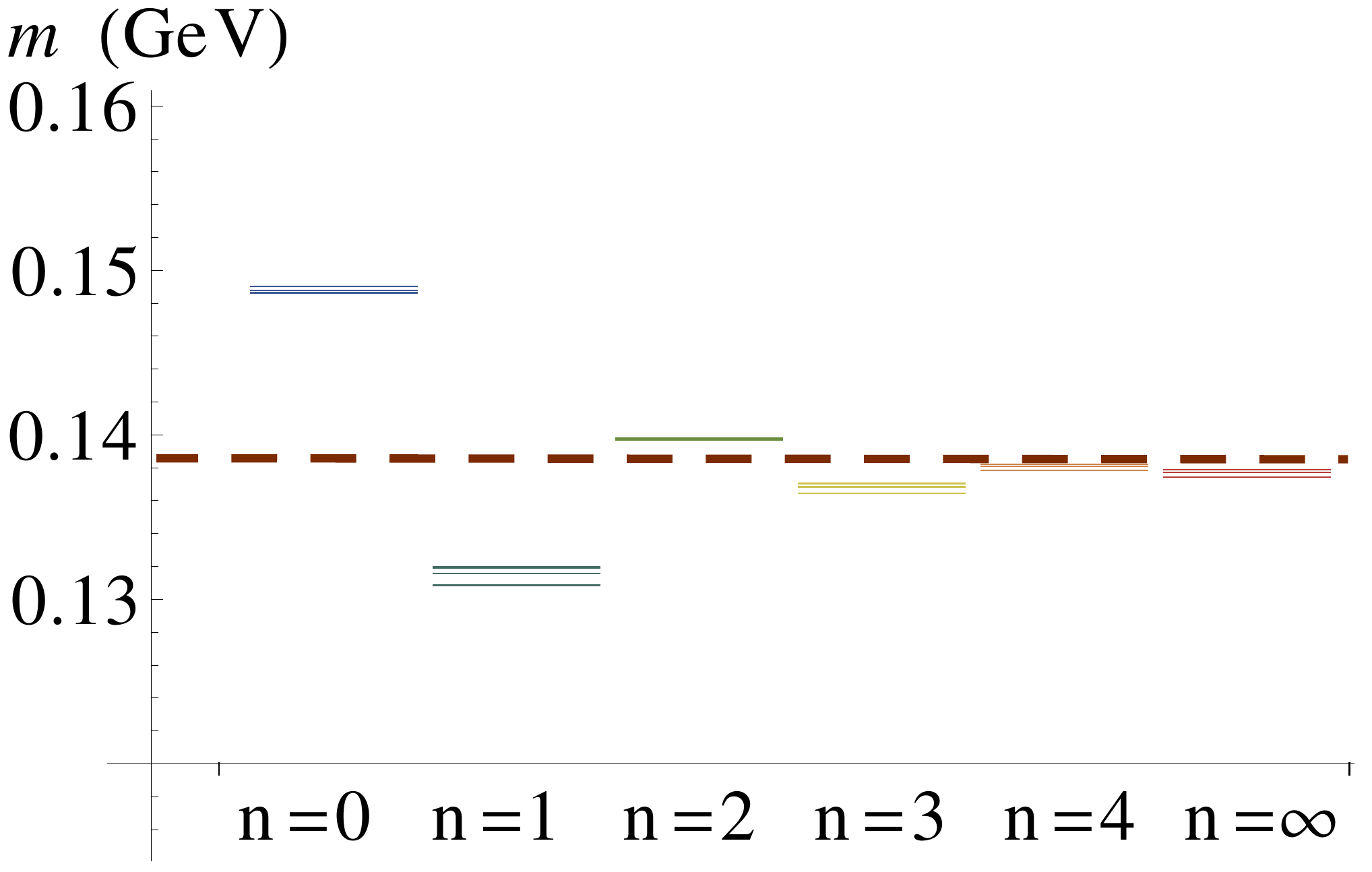}  
    \includegraphics[width=0.42\textwidth]{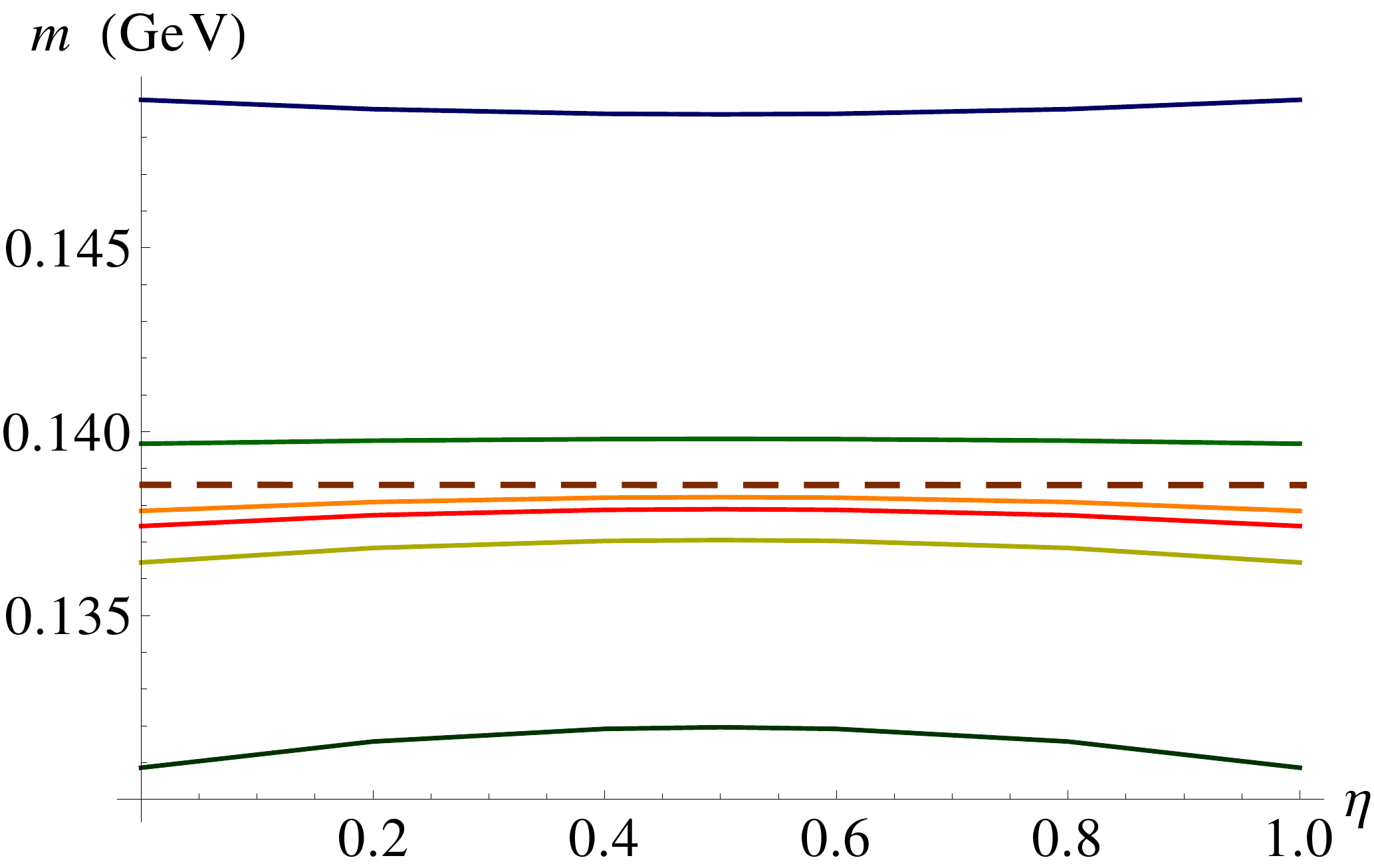} 
\caption{\emph{Left panel}: Pion mass as a function of the number of recursive steps used in the construction
of the relevant quark-gluon vertex. For each $n$, several lines are plotted for a set of representative values
of $\eta$ spanning the entire range of possible values. 
\emph{Right panel}: Pion mass plotted as a function of $\eta$; different lines
correspond to different $n$. The symmetry of the graph with respect to $\eta\leftrightarrow (1-\eta)$ characterizes
the equal-mass-constituent case. In both panels, the dashed line marks the experimental mass value for comparison.}
\label{fig:pion}       
\end{figure*}

\begin{figure}
\centering
  \includegraphics[width=0.42\textwidth]{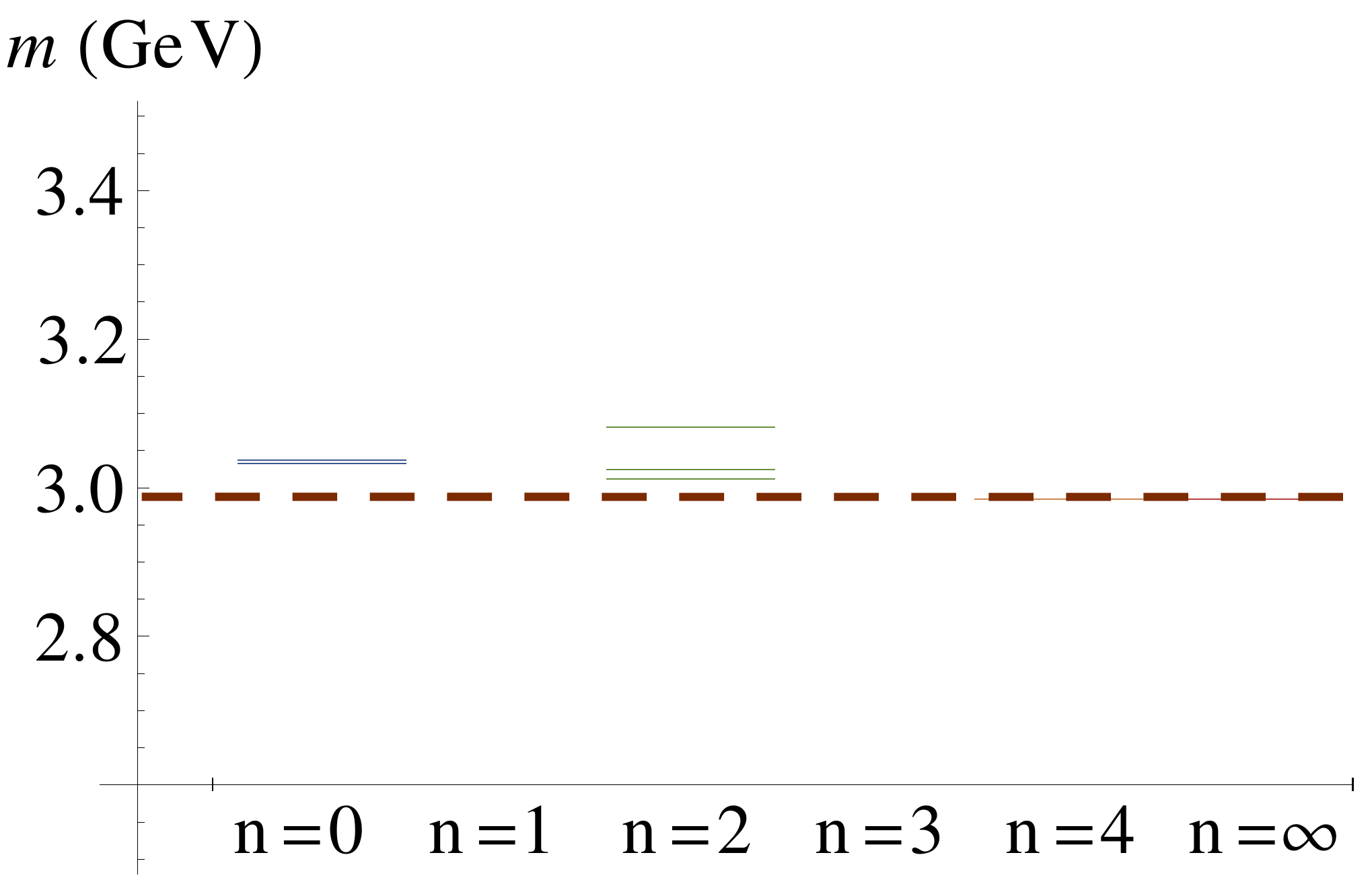}
   \includegraphics[width=0.42\textwidth]{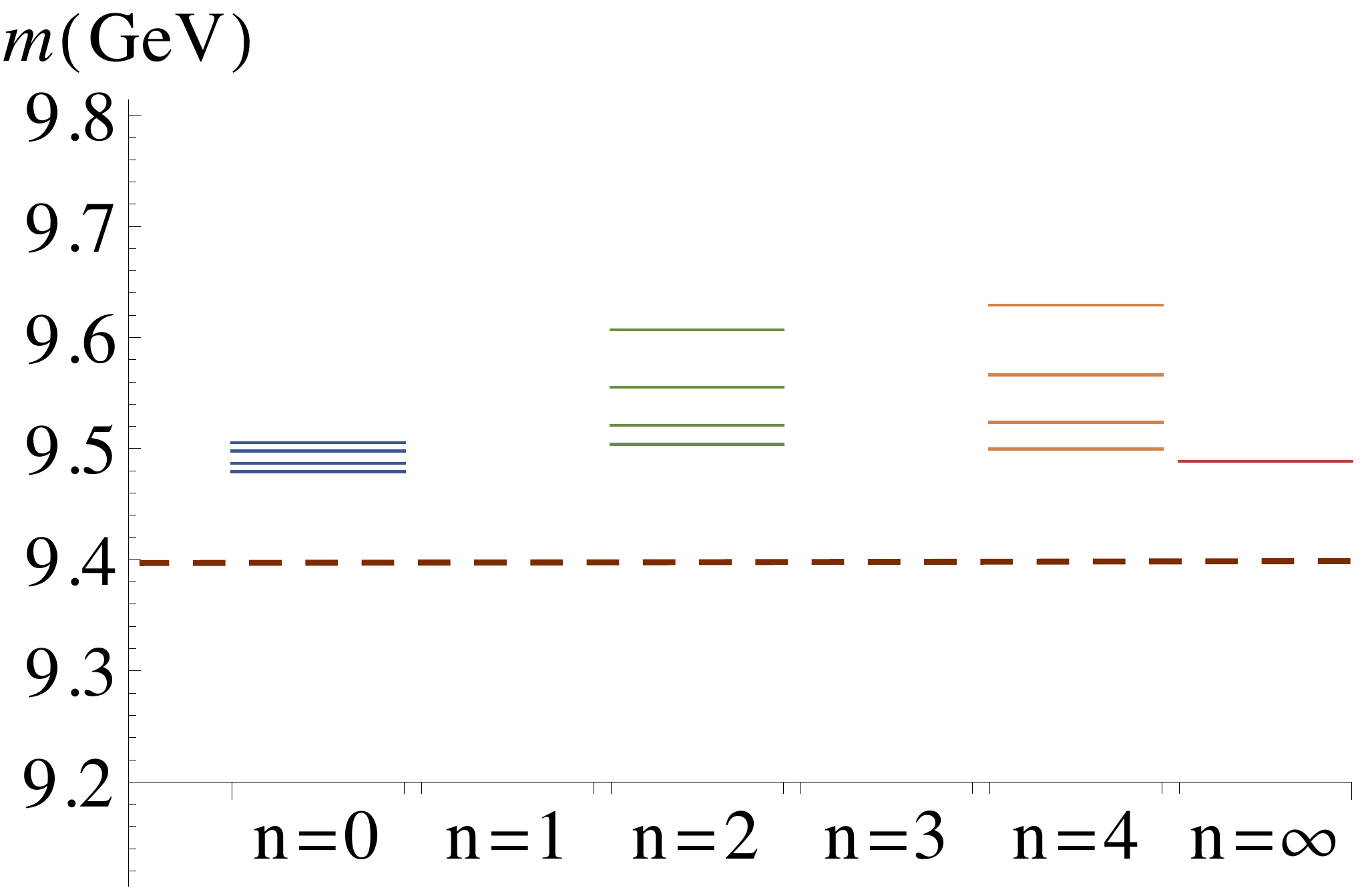}
\caption{Same as the left panel of Fig.~\ref{fig:pion} for the charmonium (left panel) and 
bottomonium (right panel) pseudoscalar ground states. Note that no lines are plotted where no solution for the BSE 
was found at a particular order $n$ or for a particular value of $\eta$. Note also that as a general pattern no solutions are found
for those situations considered here with an odd value of $n$. This qualitative difference between odd and even $n$
is not new; it has been explored and detailed in \cite{Bhagwat:2004hn}, 
and also will be further investigated in future works \cite{GomezRocha:2014mn}.}
\label{fig:quarkonia}       
\end{figure}

As a first illustration we gauge the dependence on $\eta$ for the already-known case of equal-mass quarks
in the case of the pion, as presented in Fig.~\ref{fig:pion}; this serves as a guide for the heavy-light case
in terms of a systematic error due to one particular choice for $\eta$. What we find is only a light
variation of the order of a few percent of the pion's mass for each stage in the recursive approach compared
to the fully dressed quark-gluon vertex in our scheme (see below). The same needs to be done for the 
pseudoscalar quarkonia, presented in Fig.~\ref{fig:quarkonia}: there, while the effects look larger in 
the figure, their relative order of magnitude is still at the few-percent level, comparable with the pion case.
In the figures, we also show a dashed horizontal line to indicate the respective experimental mass value for easy
reference.
Note also that not for all cases of $\eta$ a numerical solution can easily be found, in which cases the
corresponding lines in the figure are missing. More precisely, the correction to RL $\sim$ 1.6\% for $\eta=0.5$
in the charmonium case. The systematic error due to the model-artificial $\eta$-dependence can be quantified to
be smaller than 6\%. For pseudoscalar bottomonium, we obtain a correction to the RL result of $\sim$ 0.2\% for 
$\eta=0.5$, where the systematic error due to the $\eta$-dependence is smaller than 1.5\%.

The construction of the quark-gluon vertex follows a recursive pattern based on its DSE. Two correction terms to the
bare vertex, the so-called \emph{abelian} and \emph{non-abelian} gluon corrections, are considered and combined
effectively to a single term with unified form and variable dependence. The combination is weighted with an
overall factor $\mathcal{C}$, which, in turn can be used to serve as a model parameter and adjusted to 
data; in the case of \cite{Bhagwat:2004hn} this was fit to data from lattice-regularized QCD on the propagator level.
The effective equation to obtain the quark-gluon vertex reads
\begin{equation}\label{eq:qgvdse}
\Gamma_\mu^\mathcal{C}(p)=\gamma_\mu-\mathcal{C}\,\gamma_\rho\, S(p)\,\Gamma_\mu^\mathcal{C}(p)\,S(p)\,\gamma_\rho\;,
\end{equation}
which illustrates the dependence on $\mathcal{C}$. At the same time, it is clear that this equation can be 
approached with a recursive procedure and that the bare quark-gluon vertex serves as a starting value, 
$\Gamma_{\mu,0}^\mathcal{C}(p)=\gamma_\mu$. The recursion relation is
\begin{equation}\label{eq:recursion}
\Gamma_{\mu,n}^\mathcal{C}(p)=-\mathcal{C}\,\gamma_\rho\, S(p)\,\Gamma_{\mu,n-1}^\mathcal{C}(p)\,S(p)\,\gamma_\rho
\end{equation}
and the final result for the quark-gluon vertex is obtained by 
$\Gamma_\mu^\mathcal{C}(p)=\sum_{n=0}^\infty \Gamma_{\mu,n}^\mathcal{C}(p)$.

In view of these relations, additional information for the interpretation of Figs.~\ref{fig:pion} and \ref{fig:quarkonia}
is at hand: The value of $\mathcal{C}=0.51$ was chosen in \cite{Bhagwat:2004hn} and is adopted here without change together
with the values $\mathcal{G}=0.69$ GeV, $m_{u,d} = 0.01$ GeV, $m_s=0.166$ GeV, $m_c=1.33$ GeV, and $m_b=4.62$ GeV;
this enables a reasonable phenomenological description of the equal-mass pseudoscalar and vector states
and is thus an ideal choice for our study, where a quantification of systematics beyond RL truncation is sought.
Also, we can now interpret $n$ as the number of times the recursion in Eq.~(\ref{eq:recursion}) has been 
applied to the bare-vertex RL case in the quark DSE and the consistent BSE: $n=0$ corresponds to the RL truncation
itself and is thus completely insensitive to the parameter $\mathcal{C}$. $n=\infty$ is obtained by an exact summation
of all terms in the series construction of $\Gamma_\mu^\mathcal{C}(p)$ and offers a unique perspective that reaches
beyond a set of results obtained using any finite number of terms.

\section{Results and Discussion}
\label{sec:results}

After the necessary details have been laid out and checks for the equal-mass cases have been presented, we now
focus on the goal of our study, namely, to obtain information about the importance of dressings of the quark-gluon
vertex such as the one sketched above in heavy-light mesons. This is particularly interesting, since
an immediate and comprehensive phenomenologically successful sophisticated RL-study of heavy-light mesons is still missing in
the repertoire of the DSBSE approach; see \cite{Rojas:2014aka} for a recent account of efforts and status of 
current research in this direction.

\begin{figure}[t!]
\centering
  \includegraphics[width=0.42\textwidth]{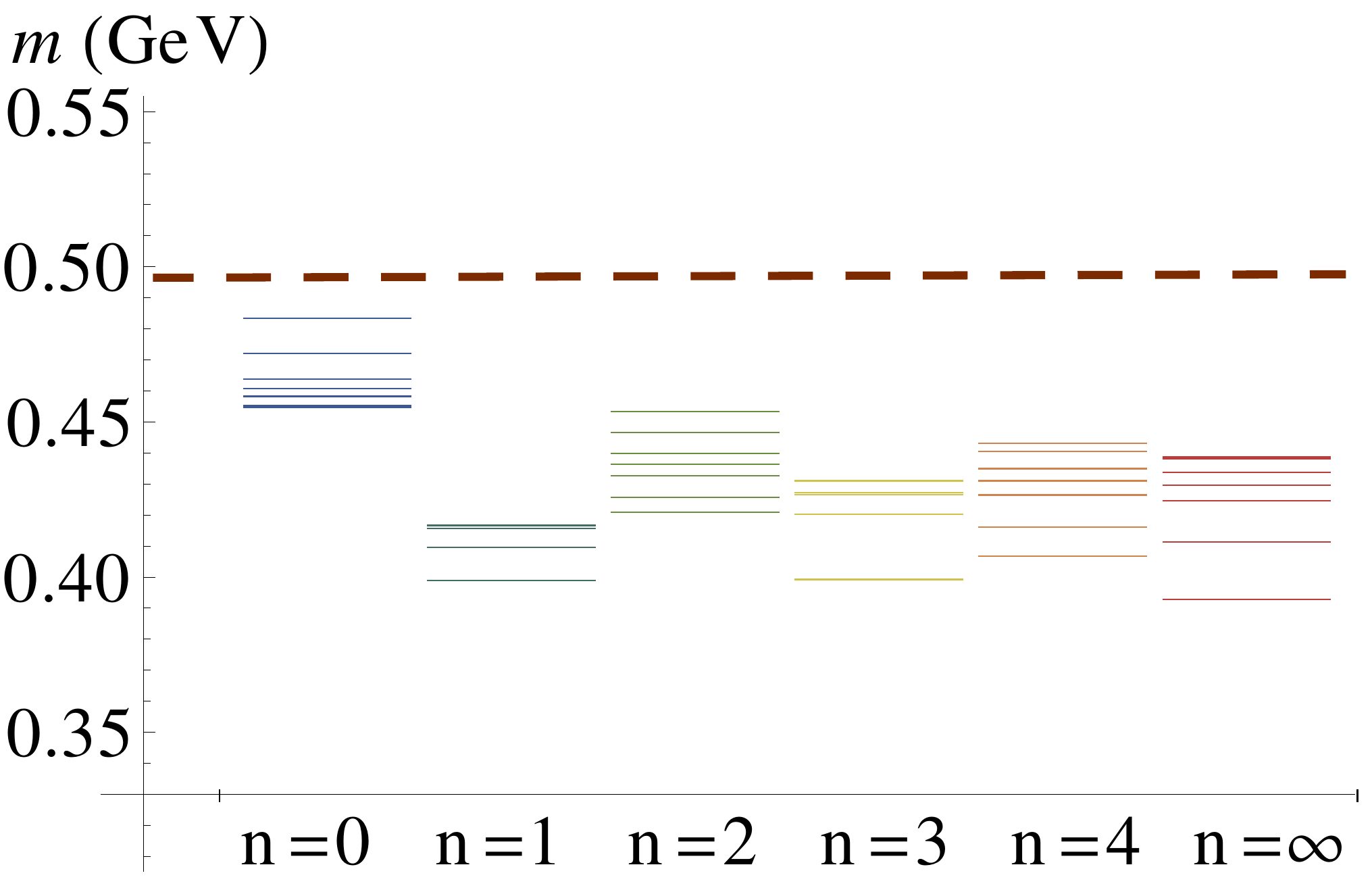}
   \includegraphics[width=0.42\textwidth]{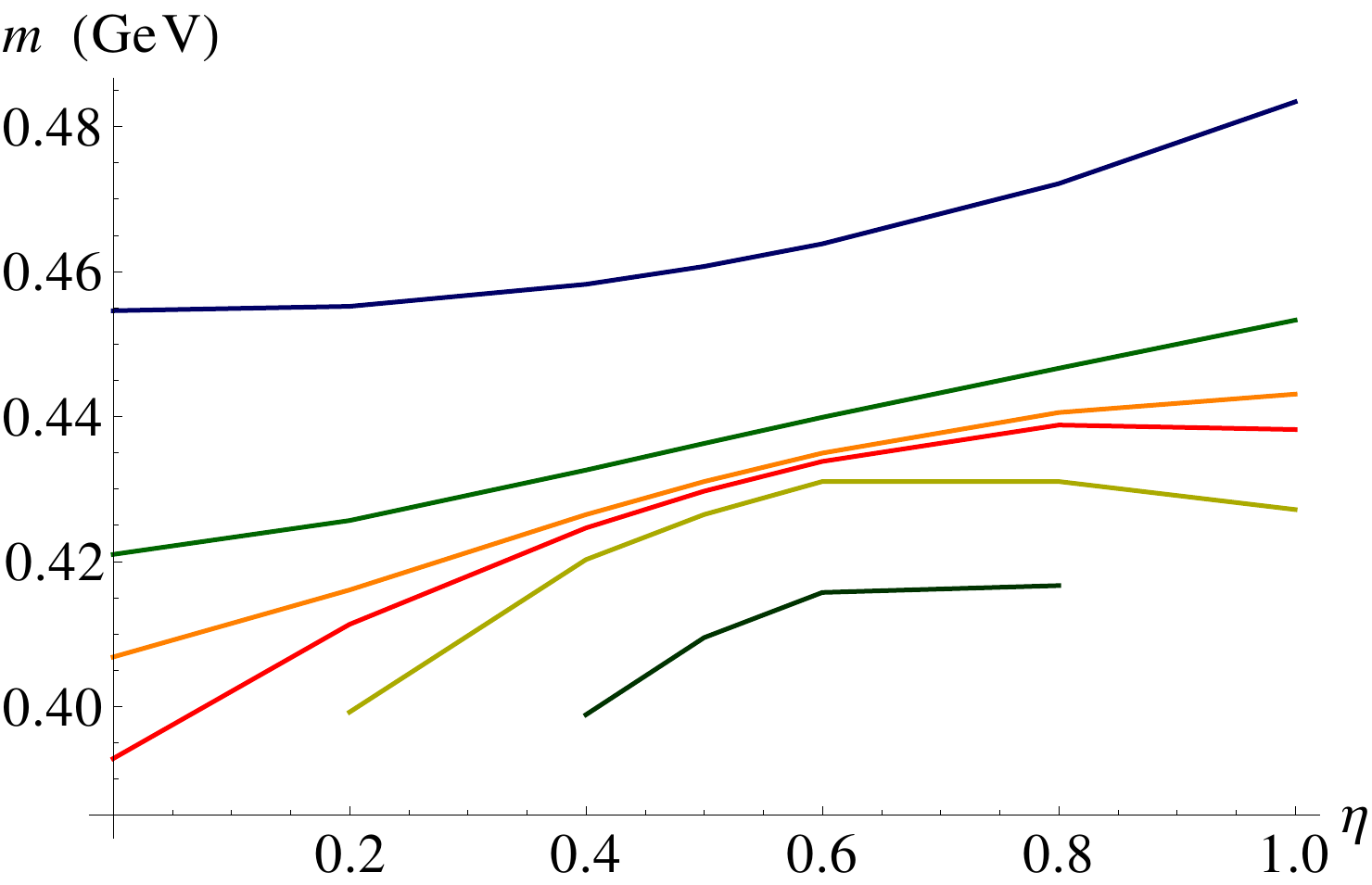}
\caption{Same as Fig.~\ref{fig:pion} for the kaon mass. The experimental value is above results and not shown in right panel.
  }
\label{fig:kaon}       
\end{figure}

As a first proving ground for our approach we investigated the kaon which is the prototype of the 
unequal-mass constituent meson in every DSBSE study. The results shown in Fig.~\ref{fig:kaon} are 
completely analogous to the ones shown for the pion in Fig.~\ref{fig:pion}. In our setup we find
a correction to RL of $\sim$ 7\% for $\eta=0.5$. The systematic error due to the $\eta$-dependence can 
be quantified to be smaller than 10\%. The symmetry with regard to $\eta\leftrightarrow (1-\eta)$
apparent for the pion in the right panel of Fig.~\ref{fig:pion} is clearly broken in the kaon
case due to the unequal masses of the quarks. We also note here that for the kaon there
are no intersections of curves in the right panel of Fig.~\ref{fig:kaon}: for each value of
$\eta$ the convergence pattern from $n=0$ towards $n=\infty$ is the same. While not shown here 
for the sake of brevity, we note that this behavior is not guaranteed by any particular aspect
of our setup and that the curves for the different $n$ can indeed intersect in other cases.

\begin{figure}[b!]
\centering
  \includegraphics[width=0.42\textwidth]{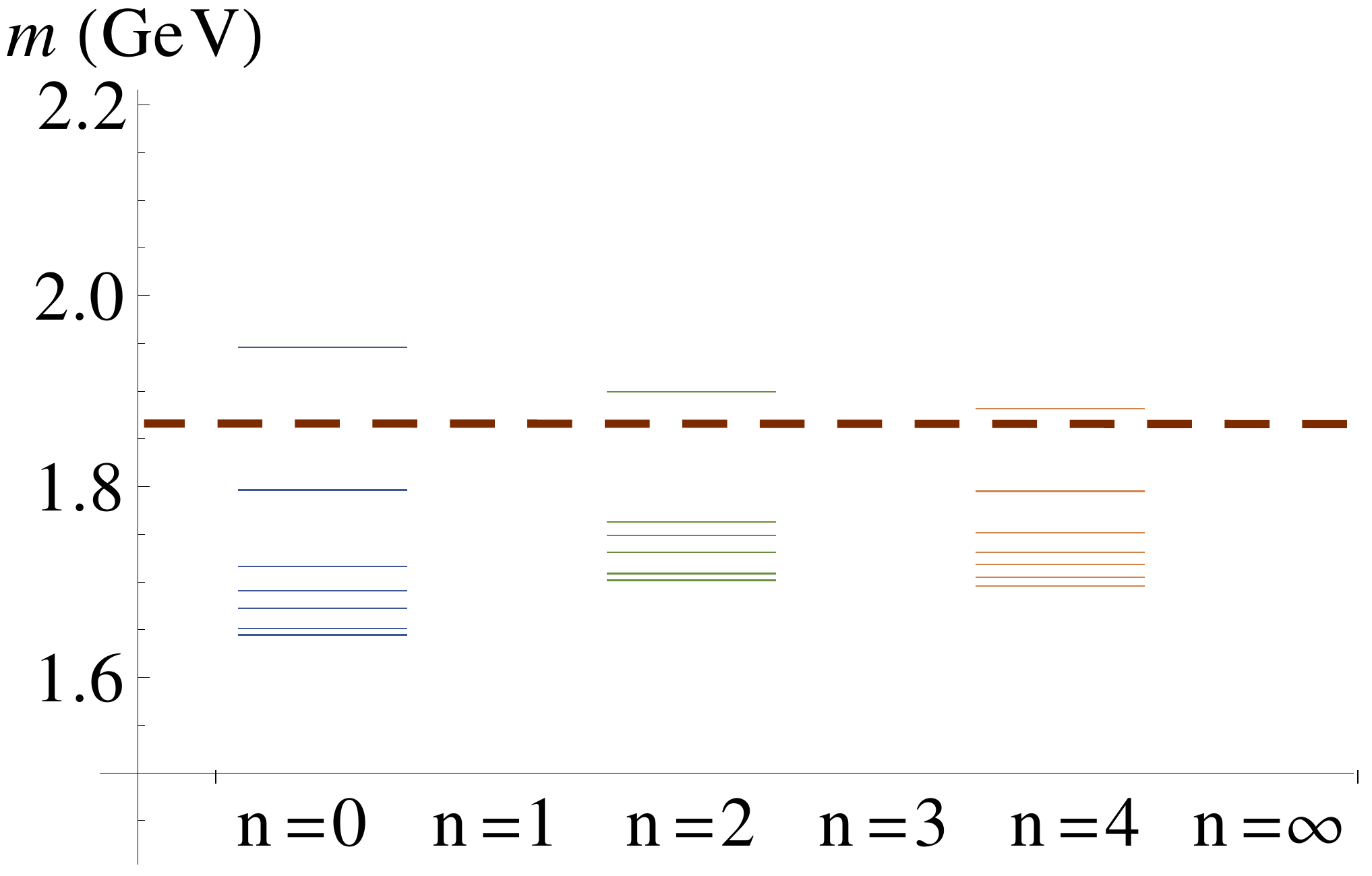}
   \includegraphics[width=0.42\textwidth]{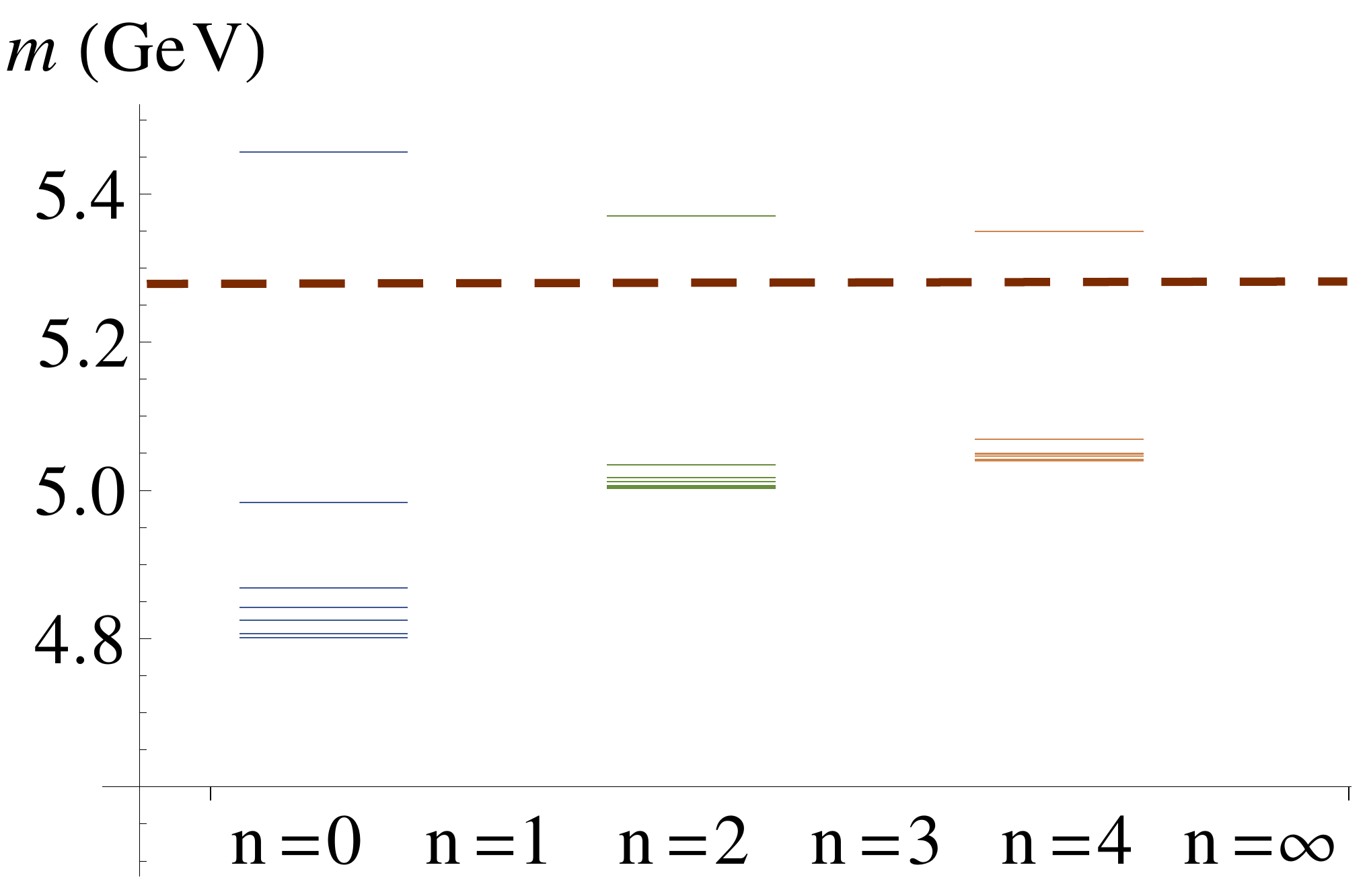}
\caption{Same as the left panel in Fig.~\ref{fig:pion} for the $D$ (left panel) and $B$ (right panel) meson masses.
}

\label{fig:heavylight}       
\end{figure}

The two heavy-light meson masses investigated here are those of the $D$ and $B$ mesons, for which
we show our results in the left and right panels of Fig.~\ref{fig:heavylight}, respectively.
In terms of quantitative information we extract the following numbers: 
For the $D$ meson we find a correction to RL truncation of $\sim$ 5\% for a choice of $\eta=0.75$.
The corresponding systematic error due to the model-artificial $\eta$-dependence can only be
weakly constrained to be less than 15\%, which somewhat blurs possible conclusions from the 
differences among results for different $n$, but is still a valuable guide for sophisticated 
studies analogous to the present. The result for the $D$ meson mass in the fully dressed case in our
scheme underestimates the experimental value by about 10\%, which is still reasonable, given 
the considerations above.

For the $B$ meson we obtain a correction to RL truncation of $\sim$ 2\% for $\eta=0.95$.  
The error due to the $\eta$-dependence is somewhat smaller than for the $D$, namely below 12\%. 
Our result for the $B$ mass in the fully-dressed case again underestimates
the experimental value, by about 4\%, but yet again lies within a reasonable range given the $\eta$-variations quoted above.
Regarding the values chosen for $\eta$ we note that they are typical values chosen in a DSBSE meson
calculation. For example, in the original treatment of the present interaction model in its RL setup in 
\cite{Munczek:1983dx}, the values chosen for $\eta$ in the $D$ and $B$ meson cases were $0.8$ and $0.92$,
respectively. We also note that the authors in \cite{Munczek:1983dx} find mass-variations with $\eta$
of similar size to our own; they quote that such variations are smaller than $15$ \%.

\section{Conclusions}

Based on previous works we investigated the $K$, $D$, and $B$ meson masses in a systematic
truncation scheme of the DSBSE system with a simplifying but far-reaching model-interaction setup.
After an estimate of systematic model-inherent errors we proceed to quantify corrections to
the popular RL truncation. In particular, a comparison of the dressing effects for the equal-mass 
and heavy-light cases could suggest that a simultaneous phenomenological description
of both kinds of pseudoscalars in an RL treatment can in fact be successful, which is
an important piece of outlook for future sophisticated hadron studies in this approach \cite{Popovici:2014pha}.
Next steps for the work presented here are the inclusion of vector mesons and a
direct check of heavy-quark symmetry predictions like they have been performed recently
in a similar setup in relativistic Hamiltonian dynamics \cite{GomezRocha:2012zd}.

\begin{acknowledgements}
We acknowledge helpful conversations with C.\,Popovici, H.\,Sanchis-Alepuz, P.~C.~Tandy, and R.~Williams. 
This work was supported by the Austrian Science Fund (FWF) under project no.\ P25121-N27.
\end{acknowledgements}


\begin{thebibliography}{36}
\providecommand{\natexlab}[1]{#1}
\providecommand{\url}[1]{\texttt{#1}}
\expandafter\ifx\csname urlstyle\endcsname\relax
  \providecommand{\doi}[1]{doi: #1}\else
  \providecommand{\doi}{doi: \begingroup \urlstyle{rm}\Url}\fi

\bibitem[Bashir et~al.(2012)Bashir, Chang, Cloet, El-Bennich, Liu,
  et~al.]{Bashir:2012fs}
A.~Bashir, L.~Chang, I.~C. Cloet, B.~El-Bennich, et~al.
\newblock \emph{Commun.Theor.Phys.}, 58:\penalty0 79--134, 2012.

\bibitem[Cloet and Roberts(2014)]{Cloet:2013jya}
Ian~C. Cloet and Craig~D. Roberts.
\newblock \emph{Prog.Part.Nucl.Phys.}, 77:\penalty0 1--69, 2014.

\bibitem[Blank and Krassnigg(2011{\natexlab{a}})]{Blank:2010bp}
M.~Blank and A.~Krassnigg.
\newblock \emph{Comput. Phys. Commun.}, 182:\penalty0 1391, 2011{\natexlab{a}}.

\bibitem[Krassnigg(2009)]{Krassnigg:2009zh}
A.~Krassnigg.
\newblock \emph{Phys. Rev. D}, 80:\penalty0 114010, 2009.

\bibitem[]{Fischer:2014xha}
C.~S.~Fischer, S.~Kubrak, and R.~Williams
\newblock  {arXiv:1406.4370}.

\bibitem[]{Alkofer:2008tt}
R.~Alkofer, C.~S.~Fischer, F.~Llanes-Estrada, and K.~Schwenzer
\newblock \emph{Ann. Phys.}, 324:\penalty0 106, 2009.

\bibitem[Maris and Roberts(1997)]{Maris:1997tm}
Pieter Maris and Craig~D. Roberts.
\newblock \emph{Phys. Rev. C}, 56:\penalty0 3369--3383, 1997.

\bibitem[Maris and Tandy(2000)]{Maris:2000sk}
Pieter Maris and Peter~C. Tandy.
\newblock \emph{Phys. Rev. C}, 62:\penalty0 055204, 2000.

\bibitem[Holl et~al.(2004)Holl, Krassnigg, and Roberts]{Holl:2004fr}
A.~Holl, A.~Krassnigg, and C.~D. Roberts.
\newblock \emph{Phys. Rev. C}, 70:\penalty0 042203(R), 2004.

\bibitem[Maris and Tandy(2006)]{Maris:2005tt}
P.~Maris and P.~C. Tandy.
\newblock \emph{Nucl. Phys. Proc. Suppl.}, 161:\penalty0 136--152, 2006.

\bibitem[Holl et~al.(2005)Holl, Krassnigg, Maris, Roberts, and
  Wright]{Holl:2005vu}
A.~Holl, A.~Krassnigg, P.~Maris, C.~D. Roberts, and S.~V. Wright.
\newblock \emph{Phys. Rev. C}, 71:\penalty0 065204, 2005.

\bibitem[Maris and Tandy(1999)]{Maris:1999nt}
Pieter Maris and Peter~C. Tandy.
\newblock \emph{Phys. Rev. C}, 60:\penalty0 055214, 1999.

\bibitem[Bhagwat and Maris(2008)]{Bhagwat:2006pu}
M.~S. Bhagwat and P.~Maris.
\newblock \emph{Phys. Rev. C}, 77:\penalty0 025203, 2008.

\bibitem[Maris(2007)]{Maris:2006ea}
Pieter Maris.
\newblock \emph{AIP Conf. Proc.}, 892:\penalty0 65--71, 2007.

\bibitem[Blank and Krassnigg(2011{\natexlab{b}})]{Blank:2011ha}
M.~Blank and A.~Krassnigg.
\newblock \emph{Phys. Rev. D}, 84:\penalty0 096014, 2011{\natexlab{b}}.

\bibitem[Eichmann et~al.(2008)Eichmann, Krassnigg, Schwinzerl, and
  Alkofer]{Eichmann:2007nn}
G.~Eichmann, A.~Krassnigg, M.~Schwinzerl, and R.~Alkofer.
\newblock \emph{Annals Phys.}, 323:\penalty0 2505--2553, 2008.

\bibitem[Eichmann et~al.(2010)Eichmann, Alkofer, Krassnigg, and
  Nicmorus]{Eichmann:2009qa}
G.~Eichmann, R.~Alkofer, A.~Krassnigg, and D.~Nicmorus.
\newblock \emph{Phys. Rev. Lett.}, 104:\penalty0 201601, 2010.

\bibitem[]{Sanchis-Alepuz:2011jn}
H.~Sanchis-Alepuz, G.~Eichmann, S.~Villalba-Chavez, and R.~Alkofer.
\newblock \emph{Phys.Rev.}, D84:\penalty0 096003, 2011.

\bibitem[Eichmann and Fischer(2013)]{Eichmann:2012mp}
Gernot Eichmann and Christian~S. Fischer.
\newblock \emph{Phys.Rev.}, D87:\penalty0 036006, 2013.

\bibitem[]{Sanchis-Alepuz:2013iia}
H.~Sanchis-Alepuz, R.~Williams, and R.~Alkofer.
\newblock \emph{Phys.Rev.}, D87:\penalty0 096015, 2013.

\bibitem[Fischer and Williams(2008)]{Fischer:2008wy}
Christian~S. Fischer and Richard Williams.
\newblock \emph{Phys. Rev. D}, 78:\penalty0 074006, 2008.

\bibitem[Fischer and Williams(2009)]{Fischer:2009jm}
Christian~S. Fischer and Richard Williams.
\newblock \emph{Phys. Rev. Lett.}, 103:\penalty0 122001, 2009.

\bibitem[Ivanov et~al.(1999)Ivanov, Kalinovsky, and Roberts]{Ivanov:1998ms}
Mikhail~A. Ivanov, Yu.~L. Kalinovsky, and Craig~D. Roberts.
\newblock \emph{Phys. Rev. D}, 60:\penalty0 034018, 1999.

\bibitem[Bender et~al.(2002)Bender, Detmold, Roberts, and
  Thomas]{Bender:2002as}
A.~Bender, W.~Detmold, C.~D. Roberts, and A.~W. Thomas.
\newblock \emph{Phys. Rev. C}, 65:\penalty0 065203, 2002.

\bibitem[Bhagwat et~al.(2004)Bhagwat, Holl, Krassnigg, Roberts, and
  Tandy]{Bhagwat:2004hn}
M.~S. Bhagwat, A.~Holl, A.~Krassnigg, C.~D. Roberts, and P.~C. Tandy.
\newblock \emph{Phys. Rev. C}, 70:\penalty0 035205, 2004.

\bibitem[Munczek and Nemirovsky(1983)]{Munczek:1983dx}
H.~J. Munczek and A.~M. Nemirovsky.
\newblock \emph{Phys. Rev. D}, 28:\penalty0 181, 1983.

\bibitem[Watson and Cassing(2004)]{Watson:2004jq}
Peter Watson and Wolfgang Cassing.
\newblock \emph{Few-Body Syst.}, 35:\penalty0 99--115, 2004.

\bibitem[Watson et~al.(2004)Watson, Cassing, and Tandy]{Watson:2004kd}
P.~Watson, W.~Cassing, and P.~C. Tandy.
\newblock \emph{Few-Body Syst.}, 35:\penalty0 129--153, 2004.

\bibitem[Fischer et~al.(2005)Fischer, Watson, and Cassing]{Fischer:2005en}
C.~S. Fischer, P.~Watson, and W.~Cassing.
\newblock \emph{Phys. Rev. D}, 72:\penalty0 094025, 2005.

\bibitem[Matevosyan et~al.(2007{\natexlab{a}})Matevosyan, Thomas, and
  Tandy]{Matevosyan:2006bk}
H.~H. Matevosyan, A.~W. Thomas, and P.~C. Tandy.
\newblock \emph{Phys. Rev. C}, 75:\penalty0 045201, 2007{\natexlab{a}}.

\bibitem[Matevosyan et~al.(2007{\natexlab{b}})Matevosyan, Thomas, and
  Tandy]{Matevosyan:2007cx}
H.~H. Matevosyan, A.~W. Thomas, and P.~C. Tandy.
\newblock \emph{J. Phys. G}, 34:\penalty0 2153--2164, 2007{\natexlab{b}}.

\bibitem[]{Williams:2014iea}
R.~Williams.
\newblock  {arXiv:1404.2545}.

\bibitem[Munczek(1995)]{Munczek:1994zz}
H.~J. Munczek.
\newblock \emph{Phys. Rev. D}, 52:\penalty0 4736--4740, 1995.

\bibitem[Chang and Roberts(2009)]{Chang:2009zb}
Lei Chang and Craig~D. Roberts.
\newblock \emph{Phys. Rev. Lett.}, 103:\penalty0 081601, 2009.

\bibitem[Heupel et~al.(2014)Heupel, Goecke, and Fischer]{Heupel:2014ina}
Walter Heupel, Tobias Goecke, and Christian~S. Fischer.
\newblock \emph{Eur.Phys.J.}, A50:\penalty0 85, 2014.

\bibitem[Gomez-Rocha et~al.()Gomez-Rocha, Hilger, and
  Krassnigg]{GomezRocha:2014mn}
M.~Gomez-Rocha, T.~Hilger, and A.~Krassnigg.
\newblock \emph{in preparation}.

\bibitem[Krassnigg and Blank(2011)]{Krassnigg:2010mh}
A.~Krassnigg and M.~Blank.
\newblock \emph{Phys. Rev. D}, 83:\penalty0 096006, 2011.

\bibitem[Blank and Krassnigg(2010)]{Blank:2010bz}
M.~Blank and A.~Krassnigg.
\newblock \emph{Phys. Rev. D}, 82:\penalty0 034006, 2010.

\bibitem[Rojas et~al.(2014)Rojas, El-Bennich, and de~Melo]{Rojas:2014aka}
E.~Rojas, B.~El-Bennich, and J.P.B.C. de~Melo.
\newblock  {arXiv:1407.3598}.

\bibitem[Popovici et~al.()Popovici, Hilger, Gomez-Rocha, and
  Krassnigg]{Popovici:2014pha}
C.~Popovici, T.~Hilger, M.~Gomez-Rocha, and A.~Krassnigg.
\newblock {arXiv:1407.7970}.

\bibitem[Gomez-Rocha and Schweiger(2012)]{GomezRocha:2012zd}
M.~Gomez-Rocha and W.~Schweiger.
\newblock \emph{Phys.Rev.}, D86:\penalty0 053010, 2012.

\end{thebibliography}
\end{document}